\documentclass[twocolumn,showpacs,preprintnumbers,amsmath,amssymb]{revtex4}

\usepackage{graphicx}
\usepackage{dcolumn}
\usepackage{bm}

\begin{document}

\title{Doping of graphene adsorbed on the a-SiO$_2$ surface}

\author{R. H. Miwa, and T. M. Schmidt}

\affiliation{Instituto de  F\'{\i}sica, Universidade  Federal de
Uberl\^andia,  C. P.  593,  38400-902, Uberl\^andia,  MG, Brazil.}

\author {W. L. Scopel}

\affiliation{Departamento de F\'{\i}sica, Universidade Federal do
  Esp\'{\i}rito Santo, Vit\'oria, 299075-910, ES-Brazil and Departamento de Ciencias
  Exatas, Universidade Federal Fluminense, 277255-250, Volta Redonda, RJ-Brazil}

\author{ A. Fazzio}

\affiliation{Instituto de F\'{\i}sica, Universidade de S\~ao Paulo, \\
    C. P. 66318, 05315-970, S\~ao Paulo, SP, Brazil.}

\date\today

\begin{abstract}

  We have  performed an {\it  ab initio} theoretical  investigation of
  graphene    sheet   adsorbed    on    amorphous   SiO$_2$    surface
  (G/a-SiO$_2$).  We  find  that  graphene adsorbs  on  the  a-SiO$_2$
  surface  through  van  der  Waals  interactions.  The  inhomogeneous
  topology  of the  a-SiO$_2$ clean  surface promotes  a  total charge
  density displacement on the  adsorbed graphene sheet, giving rise to
  electron-rich as  well as hole-rich  regions on the  graphene.  Such
  anisotropic distribution of the charge density may contribute to the
  reduction  of  the   electronic  mobility  in  G/a-SiO$_2$  systems.
  Furthermore, the adsorbed graphene sheet exhibits a net total charge
  density  gain.  In  this  case, the  graphene  sheet becomes  n-type
  doped,  however,  with  no   formation  of  chemical  bonds  at  the
  graphene--SiO$_2$  interface.  The  electronic charge  transfer from
  a-SiO$_2$  to the  graphene sheet  occurs  upon the  formation of  a
  partially occupied level lying above  the Dirac point.  We find that
  such partially occupied level  comes from the three-fold coordinated
  oxygen atoms in the a-SiO$_2$ substrate.

\end{abstract}


\maketitle




Considerable  progress has  been  made addressing  the electronic  and
structural  properties  of   graphene,  however,  the  development  of
nanodevices based on  graphene is still in its  infancy. Graphene is a
semiconductor material  with zero energy bandgap. The  valence and the
conduction  bands  are  connected  through  a  linear  energy-momentum
relation  at  the K  points  lying at  the  corners  of the  hexagonal
Brillouin zone (Dirac  points)~\cite{castronetoRMP2009}. The very high
electronic  mobility, up  to  200,000 cm$^2$/(V.s),  makes graphene  a
quite  suitable   material  to  the  development   of  new  electronic
(nano)devices.  Indeed, there are  experimental realizations  of field
effect       transistors      (FETs)      based       on      graphene
sheets~\cite{lemmeIEEE2007,liScience2008,wangPRL2008,oostingaNatMat2008,romeroACSNano2008}.
In  those  systems, SiO$_2$  has  been  used  as the  gate  dielectric
material. Lemme {\it et al.}~\cite{lemmeIEEE2007} verified a reduction
of  the  electron mobility  upon  the  graphene  interaction with  the
SiO$_2$  gate,  while  Romero  {\it  et  al.}~\cite{romeroACSNano2008}
obtained a n-type doping for  graphene in contact with a SiO$_2$ gate.
In this case, the authors attributed the presence of surface states to
the electronic charge transfer from  SiO$_2$ to the graphene sheet. In
contrast, there  are experimental  results indicating a  p-type doping
for                 graphene                 on                SiO$_2$
gate~\cite{berciaudNanoLett2009,songNanotech2010}.  Very  recently  n-
and p-type  doping of graphene,  lying on SiO$_2$ substrate,  has been
tuned    through     a    suitable    graphene--SiO$_2$    ``interface
engineering''~\cite{wangACSNano2011},  where  foreign  molecules  have
been  placed  between the  graphene  sheet  and  the SiO$_2$  surface.
Indeed,  there  are experimental  and  theoretical studies  addressing
``interface  engineering''  applied   to  modify  the  electronic  and
structural    properties    of    carbon   nanotubes    adsorbed    on
SiO$_2$~\cite{soaresNanoLett2010}   as   well    as   on   the   other
semiconductor   surfaces   like   Si(001)~\cite{OrellanaPRL2003}   and
InAs(111)~\cite{kimPRL2004}.

The  most of  the  experimental  results indicate  that  there are  no
covalent bonds  between the graphene  sheet and the  amorphous SiO$_2$
substrate   (a-SiO$_2$).  It  has   been  suggested   a  physisorption
process~\cite{stolyarovaPNAS2007,ishigamiNanoLett2007,sinitskiiACSNano2010}
where  van  der  Waals  (vdW)  interactions play  an  important  role.
Graphene sheet feels the corrugations of the a-SiO$_2$ surface, giving
rise to  electron-rich as  well as hole-rich  regions on  graphene, so
called ``electron-hole  puddles''~\cite{martinNatPhys2008}. Based upon
the aforementioned  findings, the understanding of  the electronic and
the  energetic properties,  within  an atomic  scale,  of graphene  on
a-SiO$_2$ surface is  a important issue to the  examined by using {\it
  ab initio} total energy calculations.

\begin{figure}[h]
\includegraphics[width= 7cm]{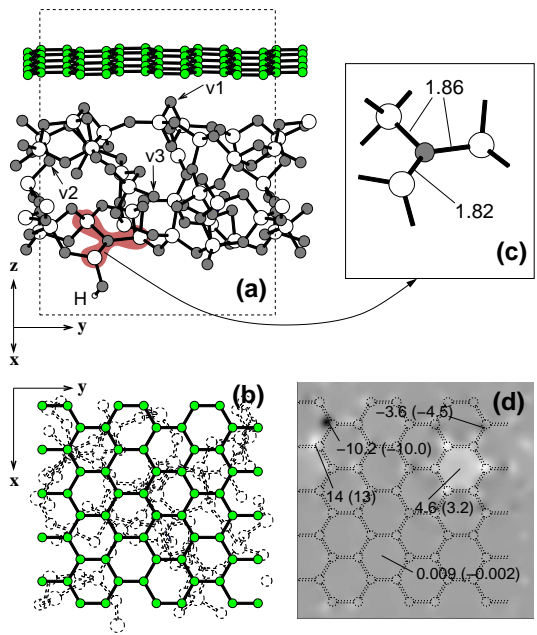}
\caption{Atomic structure of graphene adsorbed a-SiO$_2$ surface,
  G/a-SiO$_2$ (a) side-view, (b) top-view, details of the equilibrium
  geometry of O$_{\rm three-fold}$ (c).  In (a) the bottommost O
  dangling bond was saturated by a H atom and the dashed region
  indicate a periodic unit cell. In (c) the atomic distances are in
  \AA.  Filled (Empty) circles represent the O (Si) atoms. (d) Total
  charge density fluctuation ($\Delta\rho$) within a region of
  $\pm$0.5~\AA\ with respect to the (equilibrium) vertical position of
  the n-type doped graphene sheet. Within the parentheses we present
  $\Delta\rho$ for the neutral G/a-SiO$_2$ graphene sheet. $\Delta\rho$
  is in unit of $10^{14} e/{\rm cm}^2$.}
\label{structure}
\end{figure}

In this  letter we  present {\it ab  initio} calculations  of graphene
adsorbed on the a-SiO$_2$  surface, G/a-SiO$_2$. The calculations were
performed  within the  density functional  theory (DFT)  including the
long-range vdW interactions.  Our total energy results reveal that vdW
interactions   rule   the  graphene   adsorption   on  the   a-SiO$_2$
surface.  The  inhomogeneous  topology  of  a-SiO$_2$  gives  rise  to
electron-rich and hole-rich regions on the adsorbed graphene sheet, in
accordance with the  experimentally observed ``electron-hole puddles''
for  graphene lying  on  a-SiO$_2$.  However, there  is  a net  charge
density  gain on the  graphene sheet.  The electronic  charge transfer
from a-SiO$_2$ to graphene occurs  due to the formation of a partially
occupied  state above  the Dirac  point. We  find that  such partially
occupied  state   is  attributed   to  the  formation   of  three-fold
coordinated O atoms within a-SiO$_2$.


The  amorphous  structure  was   generated  through  {\it  ab  initio}
molecular  dynamics (MD)  simulations  based on  the  DFT approach  as
implemented                 in                 the                VASP
code~\cite{kressePRB1993,kresseJPhys1994,kressePRB1996}.  Initially we
obtained the  a-SiO$_2$ bulk phase. We  have used a  supercell with 96
atoms and  we made a suitable  choice of the lattice  constants of the
supercell,  12.47  and  12.96~\AA\  along  the  x  and  y  directions,
respectively,  in  order  to  match with  the  calculated  equilibrium
lattice     constant     of     an     isolated     graphene     sheet
(Fig.~\ref{structure}). While the supercell volume was minimized along
the z direction, thus, avoiding any artificial strain in the a-SiO$_2$
bulk.  We  next include a  vacuum region of $\sim$12~\AA\  parallel to
the z  direction, within the  slab method, and  then we repeat  the MD
simulations  in   order  to  get  an  a-SiO$_2$   clean  surface.   In
Ref.~\cite{scopelPRB2008}  we   present  details  on   the  generation
procedure of amorphous SiO$_2$ bulk structure. Once we obtained a well
described a-SiO$_2$  surface, we start to  investigate the equilibrium
geometry,  energetic stability,  and  the electronic  properties of  a
graphene sheet adsorbed onto  a-SiO$_2$ surface. Here the calculations
were performed by using the  DFT approach as implemented in the SIESTA
code~\cite{siesta}.  The  generalized  gradient approximation  due  to
Perdew, Burke and Ernzerhof (PBE)~\cite{PBE} was used, and the van der
Waals  interaction  was  described  within a  semiempirical  approach,
following    the   Grimme    formula~\cite{grimmeJCompChem2006}.   The
electron-ion  interactions were  calculated  by using  norm-conserving
pseudopotentials~\cite{pseudoTM}.   All the  atomic  positions of  the
G/a-SiO$_2$  system  were relaxed  by  using  the conjugated  gradient
scheme, within a force convergence criterion of 20~meV/\AA.


Figure~\ref{structure}(a)  presents  the   structural  model  for  the
physisorbed  G/a-SiO$_2$   system,  where  we   have  considered  nine
G/a-SiO$_2$   configurations  on   two  different   a-SiO$_2$  surface
structures.     At    the    equilibrium    geometry   we    find    a
graphene~--~a-SiO$_2$    equilibrium   vertical    distance   ($d_{\rm
  G-SiO_2}$) of 3.64~(3.30)~\AA\  for an a-SiO$_2$ surface corrugation
of  1.7   (1.5)~\AA~\cite{corrug}.   Although  the   relatively  short
distance between the graphene sheet and the topmost O atoms, we do not
find the  formation of C--O  chemical bonds. The strength  of graphene
adsorption  on  a-SiO$_2$  can  be  measured by  comparing  the  total
energies  of the  isolated  systems, {\it  viz.}:  graphene sheet  and
a-SiO$_2$  surface, with  the total  energy of  the  graphene adsorbed
system, G/a-SiO$_2$.  We find that  the formation of G/a-SiO$_2$ is an
exothermic   process  with   an  adsorption   energy   ($E^{ads}$)  of
6.3$\pm$0.4~{\rm  meV/\AA$^2$} Our  adsorption energy  and equilibrium
geometry  results   are  in  good  agreement   with  the  experimental
estimative of Ishigami {\it  et al.}, $E^{ads}=6~{\rm meV/\AA^2}$, and
$d_{\rm   G-SiO_2}$  =   4.2~\AA~\cite{ishigamiNanoLett2007}.   Recent
theoretical  calculations,   within  the  DFT-LDA   approach  with  no
inclusion of vdW interations, obtained $E^{ads}=1~{\rm meV/\AA^2}$ and
$d_{\rm      G-SiO_2}       =3.6~$~\AA\      for      graphene      on
a-SiO$_2$~\cite{romeroACSNano2008}.   In contrast,  other experimental
measurements obtained  $d_{\rm G-SiO_2}$ of  5.8~\AA\ for G/a-SiO$_2$,
which may indicate the presence  of foreign elements at the G--SiO$_2$
interface~\cite{songNanotech2010}.   We  next  calculate $E^{ads}$  by
turning off  the vdW contribution  from our total energy  results.  In
this case, the formation  of G/a-SiO$_2$ becomes no longer exothermic,
$E^{ads}=-0.4$~meV/\AA$^2$. Thus, showing  that graphene sheet adsorbs
on a-SiO$_2$ is mediated by vdW interactions.

Recent experimental findings indicate a corrugation of $\sim$5~\AA\ on
the     a-SiO$_2$    surface,     for    a     lateral     scale    of
$\sim$100~\AA~\cite{stolyarovaPNAS2007}.          Meanwhile         in
Ref.~\cite{ishigamiNanoLett2007}   the   authors   suggest  that   the
corrugation of the adsorbed graphene sheet should correspond to around
60\% of  the one  of the a-SiO$_2$  surface. They measured  a graphene
corrugation  of   1.9~\AA\  and  estimate  a   deformation  energy  of
1~meV/\AA$^2$.  We  obtained a very small corrugation  (up to 0.3~\AA)
and  deformation energy  (0.7~meV/\AA$^2$) for  the  adsorbed graphene
sheet, whereas our simulated  a-SiO$_2$ surface exhibits a corrugation
of $\sim$1.6~\AA. Here we would be able to describe the corrugation of
G/a-SiO$_2$ in  a suitable way by  increasing the size  of the surface
unit cell,  and performing an additional total  energy minimization of
the  surface  area.  In  this  case,  we may  find  higher  values  of
$E^{ads}$,  since the  graphene--SiO$_2$ contact  area  will increase,
however, such  increase of  $E^{ads}$ will be  limited by  the induced
strain energy upon the deformation of the graphene sheet.

It is  worth to  note that graphene  adsorbed on a-SiO$_2$  exhibits a
quite   different    picture   when   compared    with   graphene   on
$\alpha$-quartz.   In  the  latter,  due  to the  presence  of  oxygen
dangling bonds,  there is the  formation of C--O chemical  bonds (with
$E^{ads}=300~{\rm      meV/\AA^2}$)      on      the      O-terminated
surface~\cite{kangPRB2008}.  In  addition,  it  has been  suggested  a
p-type doping  of the adsorbed graphene sheet.   Whereas in a-SiO$_2$,
which also exhibits an O-rich surface [Fig.~\ref{structure}(a)], we do
not find the formation C--O chemical bonds. In our simulated annealing
process  all   the  surface   O  atoms  become   twofold  coordinated,
suppressing  the (energetically unfavorable)  O dangling  bonds.  Very
recent  DFT-LDA calculations  indicate that  the  graphene interaction
with  the (0001)  surface of  $\alpha$-quartz  is weak,  with no  C--O
chemical bonds~\cite{nguyenPRL2011}.

Figure~\ref{structure}(d) depicts the  net charge density displacement
($\Delta\rho$) on  the adsorbed  graphene sheet.  $\Delta\rho$  can be
written as,
$$
\Delta\rho = \rho[{\rm G_{SiO_2}}] - \rho[{\rm G}].
$$ Where $\rho[{\rm G_{SiO_2}}]$ and $\rho[{\rm G}]$ represent the planar
averaged  total  charge  density  of the  adsorbed  (G/a-SiO$_2$)  and
isolated  graphene  sheets,  respectively.   The  planar  average  was
performed  by considering  a region  of 0.5~\AA\  above and  below the
graphene  sheet.   The  last  term, $\rho[{\rm  G}]$,  was  calculated
keeping the same equilibrium geometry  as that of G/a-SiO$_2$. We find
graphene regions  with $\Delta\rho > 0$  as well as  $\Delta\rho < 0$,
electron-     and    hole-rich    regions,     respectively.     Thus,
Fig.~\ref{structure}(d) indicates that there is a total charge density
fluctuation on the graphene sheet, being in accordance with the recent
experimental verification of ``electron--hole puddles'' in G/a-SiO$_2$
systems~\cite{stampferAPL2007,martinNatPhys2008}.   Here we  can infer
that  the inhomogeneous  a-SiO$_2$  surface topology  rules the  total
charge  density  fluctuation of  the  adsorbed  graphene sheet.   Such
fluctuation  on  the electronic  charge  density  distribution on  the
graphene  sheet should play  an important  role on  the experimentally
verified reduction  of the electronic  mobility for graphene  lying on
a-SiO$_2$ surface~\cite{lemmeIEEE2007}.

\begin{figure}[h]
\includegraphics[width=6.5cm]{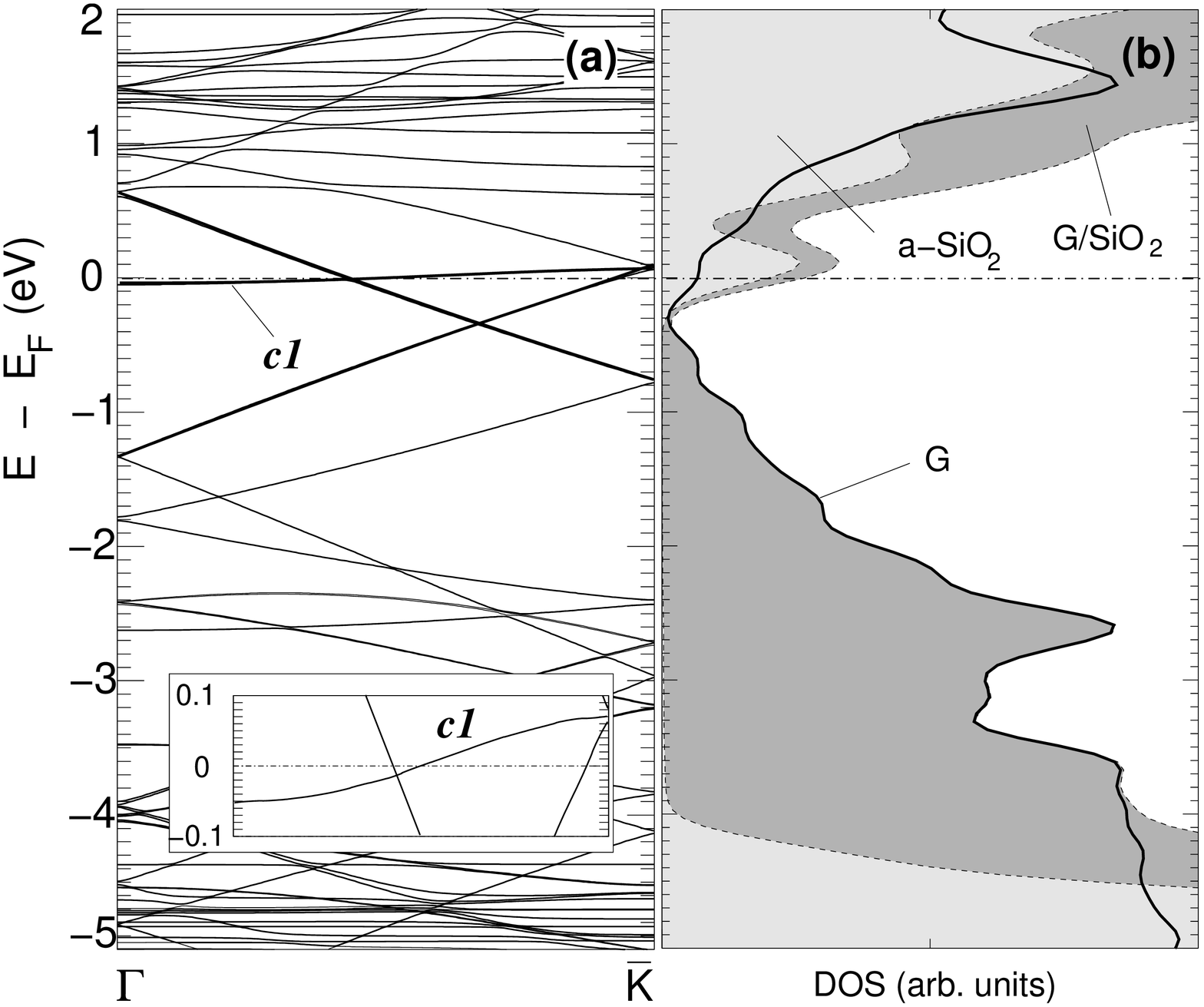}
\caption{Electronic band structure (a) and the projected density of
  states (b) of G/SiO$_2$ system. Dark shaded regions indicate the
  total density of states, and the light shaded region indicate the
  projected density of states of a-SiO$_2$ surface, and the solid line
  indicates the projected density of state of (adsorbed) graphene
  sheet. The inset in (a) present the energy dispersion of the
  partially occupied level $c1$.}
\label{pdos}
\end{figure}

There is a net electronic charge density gain of about 3~$\times
10^{13}~e/{\rm cm^2}$, Fig.~\ref{structure}(d), thus, indicating a
n-type doping on the adsorbed graphene sheet.  Our electronic band
structure calculation, Fig.~\ref{pdos}(a), confirms statement above.
The shape of the Dirac cone has been maintained, since the graphene
sheet weakly interacts with the a-SiO$_2$ substrate.  Due to the
electronic charge transfer from a-SiO$_2$ to the graphene sheet, the
Dirac point lies below the Fermi energy ($\rm E_F$). The n-type doping
occurs due to the presence of a partially occupied state ($c1$) below
the conduction band minimum (CBM) pinning the Fermi level above the
Dirac point, see inset of Fig.~\ref{pdos}(a).  Figure~\ref{pdos}(b)
presents the density of states (DOS) of G/a-SiO$_2$ (gray region), and
separately the contribution from the adsorbed graphene sheet (solid
line), and the a-SiO$_2$ surface (light gray region). We find, in
contrast with previous investigations~\cite{romeroACSNano2008}, that
$c1$ comes from the formation of three-fold coordinated O atoms
(O$_{\rm three-fold}$) in a-SiO$_2$. Those O$_{\rm three-fold}$ sites
may occur when we have a slightly hight local concentration of Si
atoms or nearby oxygen vacancies in
a-SiO$_2$~\cite{luPRL2002,sokolovNonCrystSol1997,winerPRB1988,feustonJChemPhys1989}.
The Si--O$_{\rm three-fold}$ bond lengths (1.82--1.86~\AA) are
stretched in comparison with the ones of the two fold coordinated O
atoms, $\sim$1.66~\AA, Fig.~\ref{structure}(c).  Those larger bond
lengths reduce the binding energy of the electronic states along
Si--O$_{\rm three-fold}$, giving rise to the partially occupied state,
$c1$, just below the CBM and pinning the Fermi level. We find that the
electronic states of $c1$ are localized on the Si atoms nearby O$_{\rm
  three-fold}$.  In order to verify the adequacy of the proposed
origin of $c1$, we calculate the electronic structure of G/a-SiO$_2$
by suppressing the presence of O$_{\rm three-fold}$, {\it i.  e.} we
have only two fold coordinated O atoms within the a-SiO$_2$ slab.  In
this case, we do not find the formation of $c1$, and the Dirac point
lies on the Fermi level.  That is, the adsorbed graphene sheet becomes
neutral.  Those results support the proposed role of O$_{\rm
  three-fold}$ on the non-covalent n-type doping of graphene on
a-SiO$_2$ substrate.  It is worth to note that, for the neutral
graphene sheet, we find the same picture for the electron- and
hole-rich regions, as depicted in Fig.~\ref{structure}(d), however
with different values of $\Delta\rho$.  The total charge density
fluctuation on the neutral graphene is about 25\% smaller in
comparison with the (n-type) doped ones.


 Finally, we have considered  the presence of oxygen vacancies
  (O$_{\sf V}$) in a-SiO$_2$. We have examined three plausible O$_{\sf V}$
configurations, on the topmost surface site, and on the subsurface
sites, indicated as {\sf V1}--{\sf V3} in Fig.~\ref{structure}(a),
respectively. We verify that those vacancies do not give rise to a
donor level on the a-SiO$_2$ surface. For {\sf V1} and {\sf V2} we
find the formation of Si-dimers on the surface (Si--Si bond length of
2.24~\AA) inducing the formation of occupied (empty) electronic states
near the valence (conduction) band maximum (minimum). Whereas {\sf V3}
gives rise to a deep occupied state within the a-SiO$_2$ bandgap. 



In conclusion, we find that vdW interactions rules the graphene
adsorption on the a-SiO$_2$ surface.  Through a charge density map on
the adsorbed graphene sheet we verify that the inhomogeneous a-SiO$_2$
surface topology promotes an electronic charge density displacement on
the graphene, giving rise to the experimentally verified
``electron-hole puddles''~\cite{stampferAPL2007,martinNatPhys2008}.
Such electron-rich and hole-rich regions should contribute to the
experimentally verified reduction on the electronic mobility of
graphene sheets lying on a-SiO$_2$ surfaces.  We find a net charge
density gain on the adsorbed graphene sheet, thus, characterizing a
n-type non-covalent doping on the graphene.  The electronic charge
transfer from a-SiO$_2$ to the graphene sheet comes from the formation
of a partially occupied state above the Dirac point, which is
attributed to the formation of three-fold coordinated oxygen atoms in
a-SiO$_2$.  Those three-fold coordinated oxygen atoms may occurs upon
slightly hight local concentration of Si atoms or nearby oxygen
vacancies.




\acknowledgments

The authors thanks Dr. Renato Pontes for fruitful discussions, the
financial support from the Brazilian agencies CNPq, CAPES, FAPEMIG,
FAPESP, CNPq/INCT and the computational support from CENAPAD/SP.

\bibliography{/home/hiroki/Trab/RHMiwa}

\begin{thebibliography}{32}
\bibitem[{cor()}]{corrug}
\bibinfo{note}{The a-SiO$_2$ surface corrugations was inferred by the
  calculation of the variance of vertical positions of the topmost O atoms,
  while for the adsorbed graphene sheet, we considered the variance of vertical
  positions of the graphene C atoms.}
\end{thebibliography}

\end{document}